\documentclass[10pt]{article}
\usepackage{graphicx}
\usepackage{atlasphysics} 
\usepackage{subfigure}
\usepackage{lineno}

\def\Title#1{\begin{center} {\Large #1 } \end{center}}
\def\Author#1{\begin{center}{ \sc #1} \end{center}}
\def\Address#1{\begin{center}{ \it #1} \end{center}}

\newcommand\pubblock{\rightline{\begin{tabular}{l} Proceedings of the Second Annual LHCP\\ \pubnumber\\
         \pubdate  \end{tabular}}}

\newenvironment{Abstract}{\begin{quotation} \begin{center} 
             \large ABSTRACT \end{center}\bigskip 
      \begin{center}\begin{large}}{\end{large}\end{center} \end{quotation}}

\newenvironment{Presented}{\begin{quotation} \begin{center} 
             PRESENTED AT\end{center}\bigskip 
      \begin{center}\begin{large}}{\end{large}\end{center} \end{quotation}}

\def\beq{\begin{equation}}
\def\eeq#1{\label{#1}\end{equation}}
\def\eeqn{\end{equation}}

\def\beqa{\begin{eqnarray}}
\def\eeqa#1{\label{#1}\end{eqnarray}}
\def\eeqan{\end{eqnarray}}

\def\sst{\scriptscriptstyle}

\let\bar=\overbar

\def\Dslash{\not{\hbox{\kern-4pt $D$}}}
\def\dslash{\not{\hbox{\kern-2pt $\del$}}}

\def\mt{m_t}

\def\msb{{\bar{\ssstyle M \kern -1pt S}}}

\usepackage{color}

\newcommand\pubnumber{ ATL-UPGRADE-PROC-2014-003 }

\newcommand\pubdate{\today}

\def\affiliation{
Center for High Energy Physics\\
University of Oregon, Eugene, OR 97403, U. S. A.\\
\vspace{0.3in}
On behalf of the ATLAS Collaboration, \\
}

\begin{document}
\large
\begin{titlepage}
\pubblock

\vfill
\Title{  ATLAS upgrades for the next decades  }
\vfill

\Author{ Walter Hopkins  }
\vspace{-0.1in}
\Address{\affiliation}
\vfill
\begin{Abstract}
 After the successful LHC operation at the center-of-mass energies of 7
        and 8 TeV in 2010-2012, plans are actively advancing for a series of
        upgrades of the accelerator, culminating roughly ten years from now in
        the high-luminosity LHC (HL-LHC) project, delivering of the order of
        five times the LHC nominal instantaneous luminosity along with
        luminosity leveling. The final goal is to extend the dataset from
        about few hundred \ifb\ to 3000 \ifb\ by around
        2035 for ATLAS and CMS. In parallel, the experiments need to be kept
        lockstep with the accelerator to accommodate running beyond the
        nominal luminosity this decade. Current planning in ATLAS envisions
        significant upgrades to the detector during the consolidation of the
        LHC to reach full LHC energy and further upgrades. The challenge of
        coping with the HL-LHC instantaneous and integrated luminosity, along
        with the associated radiation levels, requires further major changes
        to the ATLAS detector. The designs are developing rapidly for a new
        all-silicon tracker, significant upgrades of the calorimeter and muon
        systems, as well as improved triggers and data acquisition. This
        report summarizes various improvements to the ATLAS detector required
        to cope with the anticipated evolution of the LHC luminosity during
        this decade and the next.

\end{Abstract}
\vfill

\begin{Presented}
The Second Annual Conference\\
 on Large Hadron Collider Physics \\
Columbia University, New York, U.S.A \\ 
June 2-7, 2014
\end{Presented}
\vfill
\end{titlepage}
\def\thefootnote{\fnsymbol{footnote}}
\setcounter{footnote}{0}
%

\normalsize 


\section{Introduction}
The LHC will undergo several iterations of improvement starting with the increase of center-of-mass energy to 13 TeV in 2015. Further upgrades will increase instantaneous luminosities from $\sim1\times10^{34}$ cm$^{-2}$s$^{-1}$ to $\sim2\times10^{34}$ cm$^{-2}$s$^{-1}$ during LS2 and from $\sim2\times10^{34}$ cm$^{-2}$s$^{-1}$ to $\sim5\times10^{34}$ cm$^{-2}$s$^{-1}$ during LS3 (Fig.~\ref{fig:lhcSchedule}). \par 

These increases in luminosity allow increased sensitivity to low cross section processes such as $\ttbar H$ production. Precision measurements of Higgs couplings could yield the first indications of Beyond the Standard Model Physics. Direct searches for new physics, such as the search for the supersymmetric top partner, will also benefit from the increase in luminosity (Fig.~\ref{fig:higgStop}). The current Run 2 limits on the stop quark mass at low LSP mass are at $\sim700$ GeV. It should be possible to discover (exclude) stop quarks with masses up to 1.2 (1.0) TeV with 300 \ifb\ and up to 1.2 (1.4) TeV with 3000 \ifb. \par

The upgrades, however, come with great challenges to maintain similar performance under much harsher environments. Without proper upgrades to the ATLAS detector~\cite{atlasDet} much of the produced luminosity would be lost due to trigger limitations. The focus of the Phase-I and II upgrades is to have similar performance in Run 2 and Run 3 as during Run 1 with significantly increased instantaneous luminosities. 

\begin{figure}[tb]
\begin{center}
\subfigure{\includegraphics[width=0.65\textwidth]{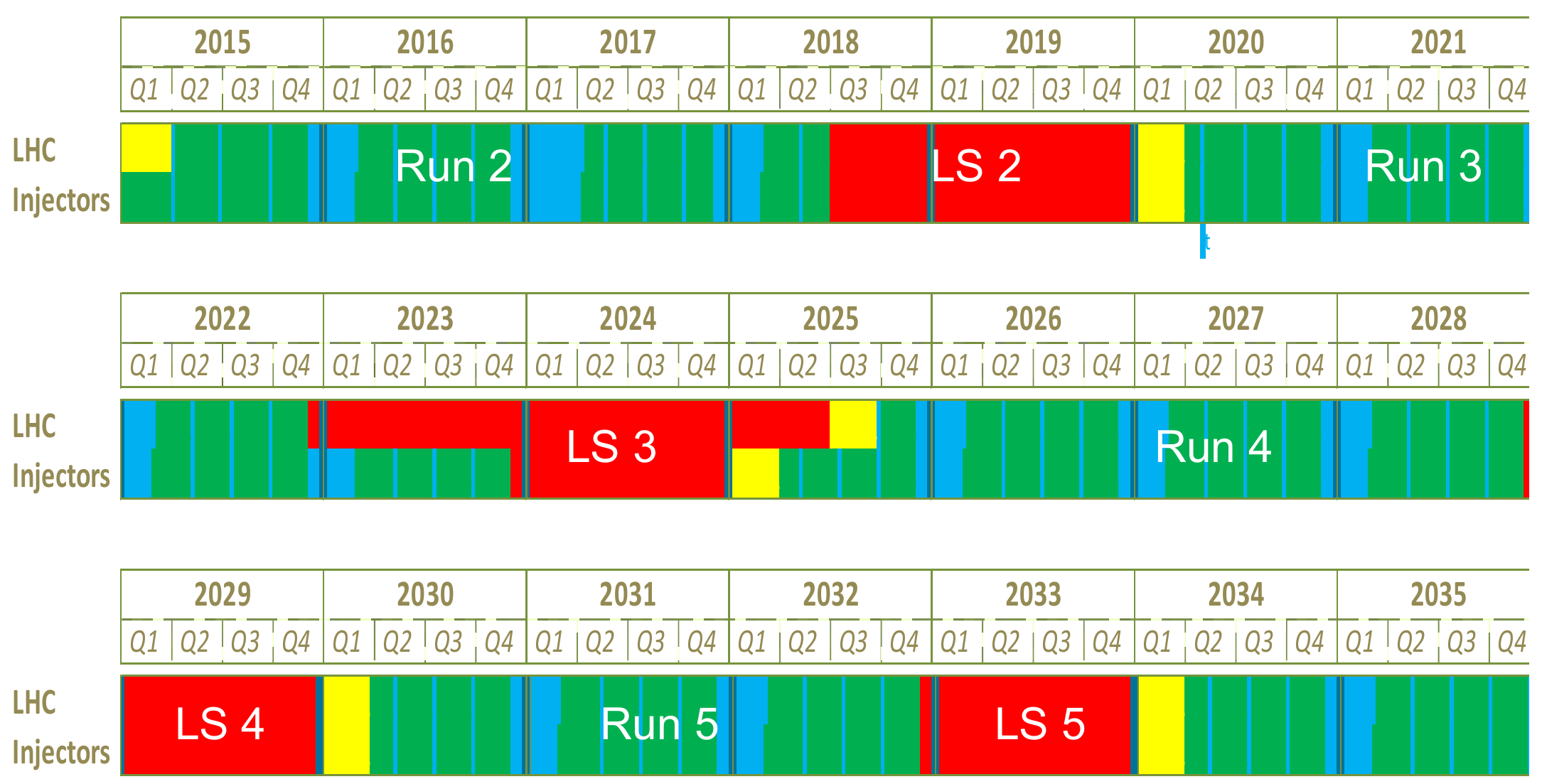}}
\end{center}
\caption{LHC schedule for the next 20 years with the status of the LHC labeled for the top colored row and the injector status on the bottom colored row. The Long Shutdowns (LS1, LS2, and LS3) will be from 2013 to 2015, mid-2018 to 2019, and 2023 to 2025, respectively. The Phase-0, I, and II upgrades are scheduled for LS1, LS2, and LS3, respectively.}
\label{fig:lhcSchedule}
\end{figure}
\begin{figure}[tb]
\begin{center}
\subfigure{\includegraphics[width=0.25\textwidth]{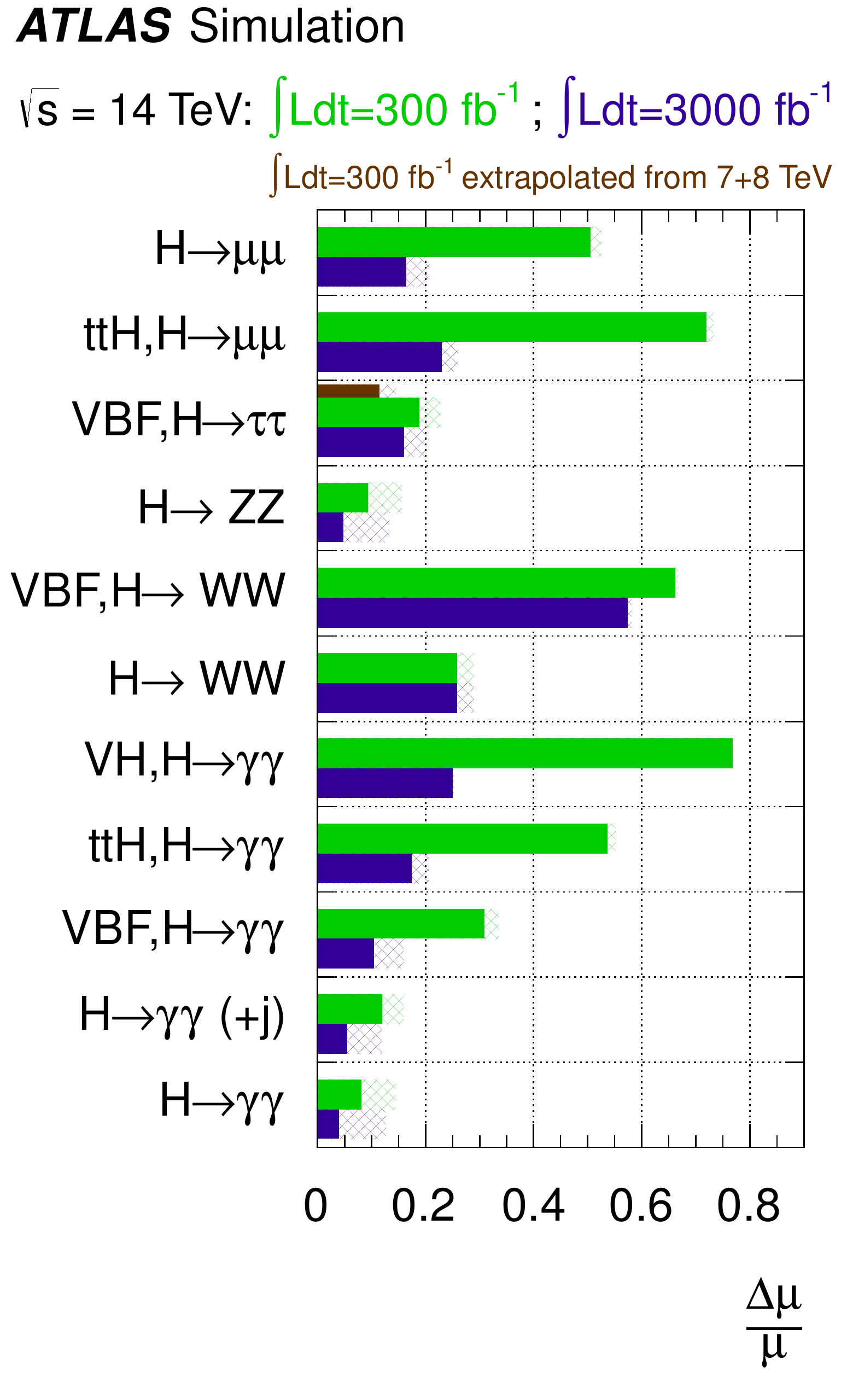}}
\subfigure{\includegraphics[width=0.55\textwidth]{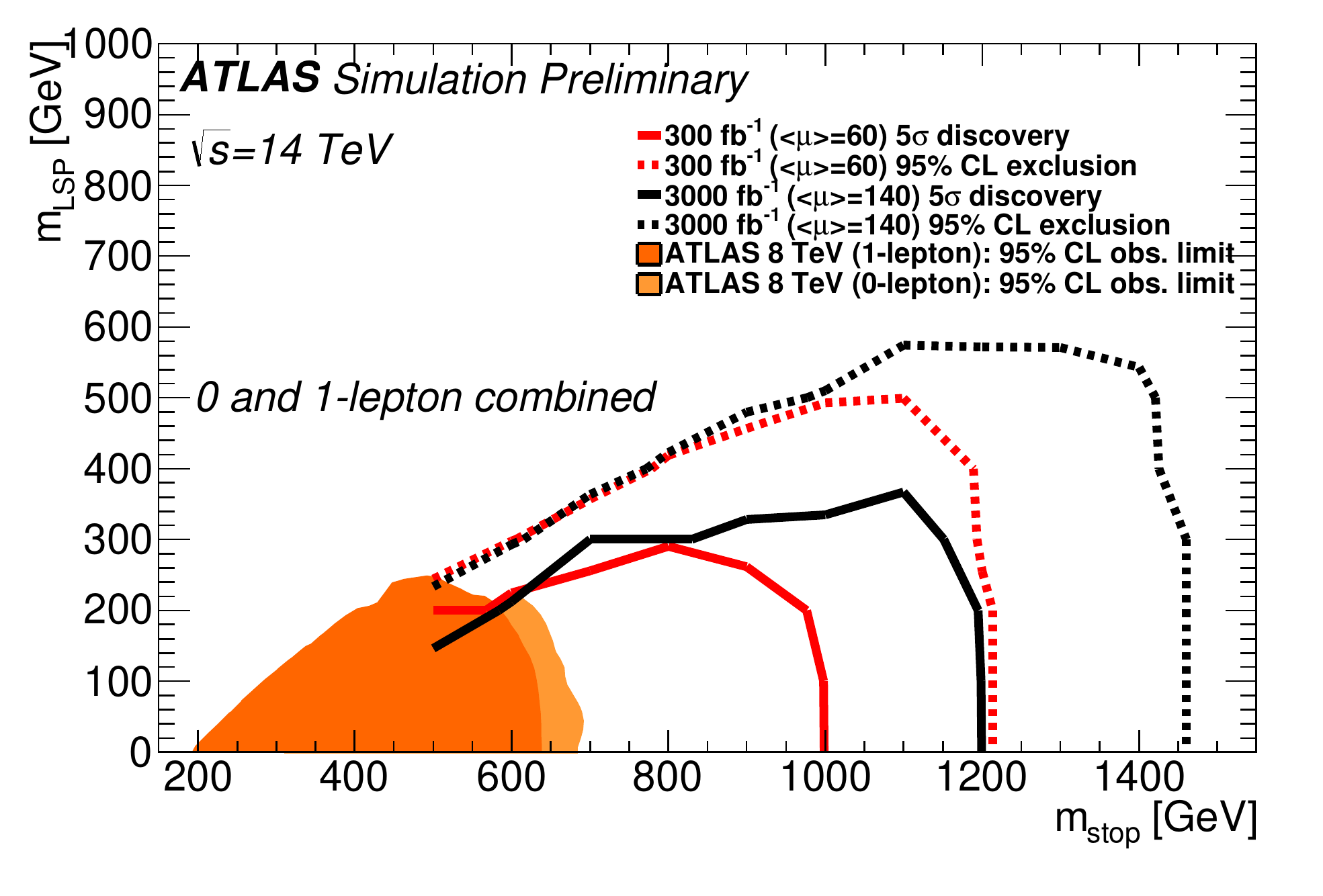}}
\end{center}
\caption{Expected precision on the measurement of signal strengths for various Higgs decay modes (left) and exclusion and discovery reach for a stop quark (right) using 300 \ifb\ and 3000 \ifb. The stop limits are shown as a function of the stop mass and the lightest supersymmetric particle (LSP) mass. }
\label{fig:higgStop}
\end{figure}

\section{Phase-0 upgrade}
The Phase-0 upgrade is occurring during the 2013-2015 shutdown and aims at achieving the design center-of-mass energy collisions of 13 TeV with 25 ns bunch crossing and instantaneous luminosities of $\sim1\times10^{34}$ cm$^{-2}$s$^{-1}$. The centerpiece of this upgrade is the insertable B-layer (IBL). This layer is a $4^{\textnormal{th}}$ silicon tracker module that is installed directly on a new, smaller, aluminum beam pipe with a radius of 25 mm (Fig~\ref{fig:ibl}). The module has a pseudorapidity coverage up to $\eta<3$ and features 3D and planar sensor design will greatly improve tracking, vertex, and b-jet identification~\cite{iblTDR}. \par

In addition to the IBL installation, the muon system will be completed along with the installation of updated LAr low voltage power supplies. 
\begin{figure}[tb]
\begin{center}
\subfigure{\includegraphics[width=0.44\textwidth]{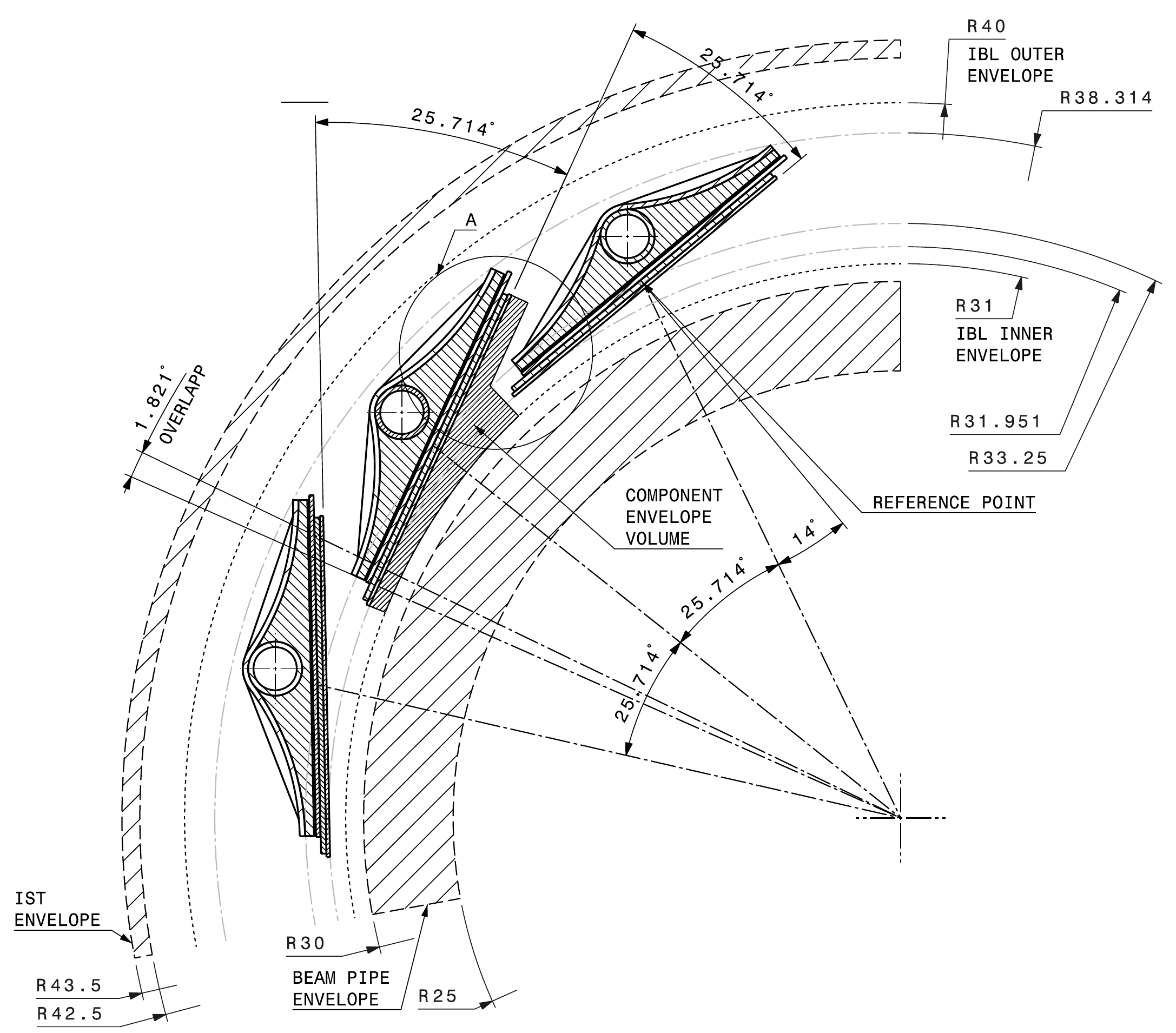}}
\subfigure{\includegraphics[width=0.37\textwidth]{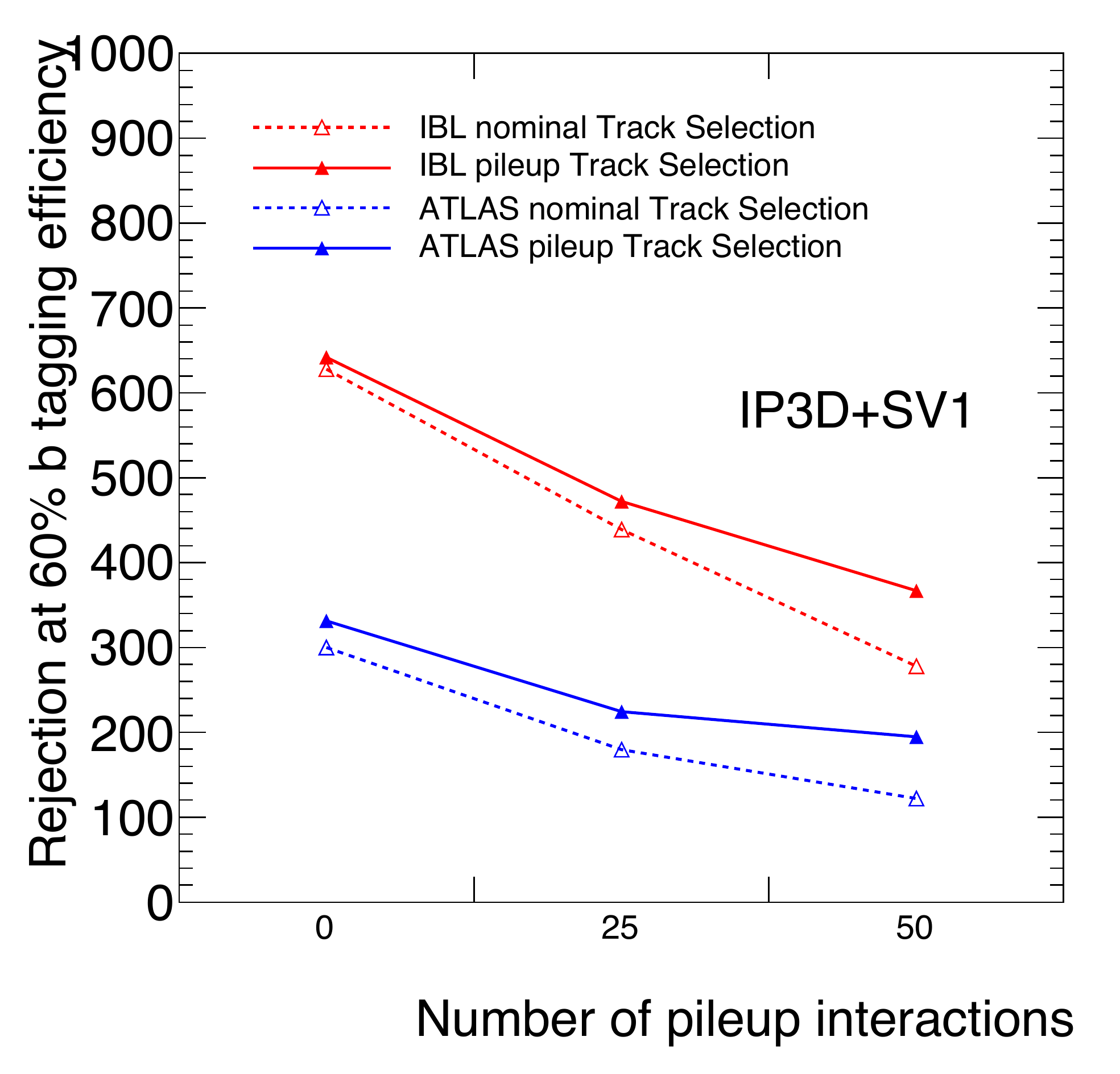}}
\end{center}
\caption{Left: IBL configuration around new beam pipe. Right: Rejection of light jets at a 60\% b-jet efficiency for \ttbar\ events with the IP3D+SV1 b-tagger.}
\label{fig:ibl}
\end{figure}

\section{Phase-I upgrade}
The major LHC upgrade for Phase-I in 2018-2019 will be a doubling of the instantaneous luminosity to $\sim2\times10^{34}$ cm$^{-2}$s$^{-1}$.
The Phase-I upgrade includes significant improvements to the muon system by the addition of the muon small wheel. The inner most wheel of the muon endcap will be replaced with a new wheel that features Thin Gap Chambers (TGC) and Micro-MEsh GAseous Structures (Micro-MEGAs) technologies~\cite{muSmallWheelTDR}. This updated wheel will prevent the drastic increase of fake muons in the forward region which would exceed the trigger bandwidth. This is accomplished with the use of high quality pointing information. The location of the new wheel as well as the rate of current Level-1 trigger muons as a function of $\eta$ are shown in Fig.~\ref{fig:smallWheel}.\par 

In addition to the muon small wheel upgrade a new fast tracker~\cite{ftkTDR} at the input of the Level-2 trigger will be implemented as well as a LAr calorimeter electronics upgrade~\cite{larTDR}. The track occupancy will drastically increase with the higher instantaneous luminosities. A fast track trigger would be included for Phase-I allowing for fast track finding at the hardware level trigger. This frees up resources for the Level-2 trigger to use more advance rejection methods.\par

The current LAr system performs an analog sum of calocell energies into $0.1\times0.1$ trigger towers with no layer information. The upgraded system will use digital sums to create supercells out of calocells with varying fine granularity (Fig.~\ref{fig:LArSCTT}). Layer information is retained and can be used by the trigger to, for example, differentiate between electrons and jets.\par

Finally, to accommodate the various changes, the trigger data acquisition system (TDAQ) must also be upgraded~\cite{tdaqTDR}. The new small wheel readout will be included in the Level-1 muon endcap trigger, lowering forward trigger rates significantly, while new Level-1 feature extraction processors will be included to take advantage of the finer granularity LAr calorimeter readout. 
       
\begin{figure}[tb]
\begin{center}
\subfigure{\includegraphics[width=0.49\textwidth]{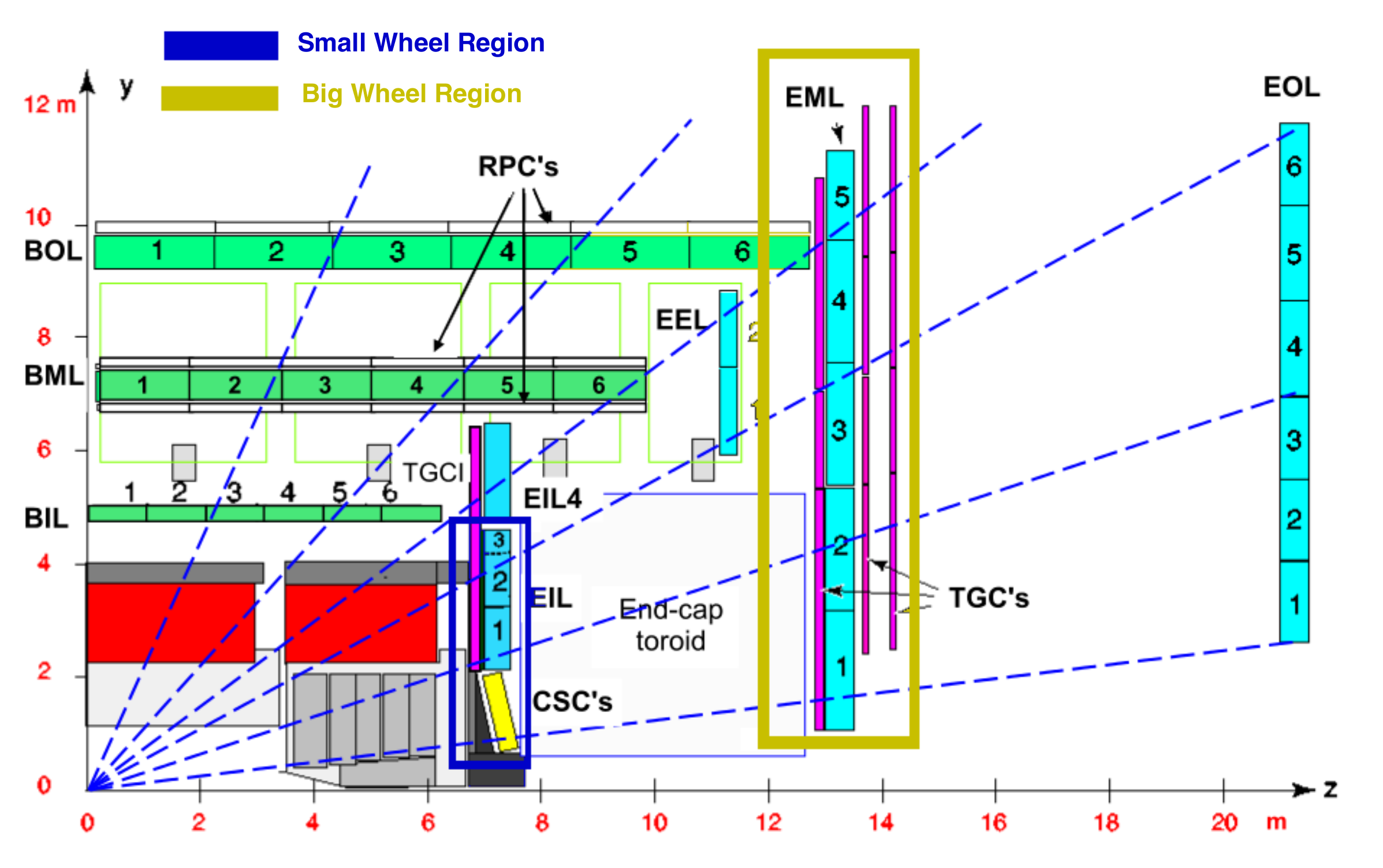}}
\subfigure{\includegraphics[width=0.4\textwidth]{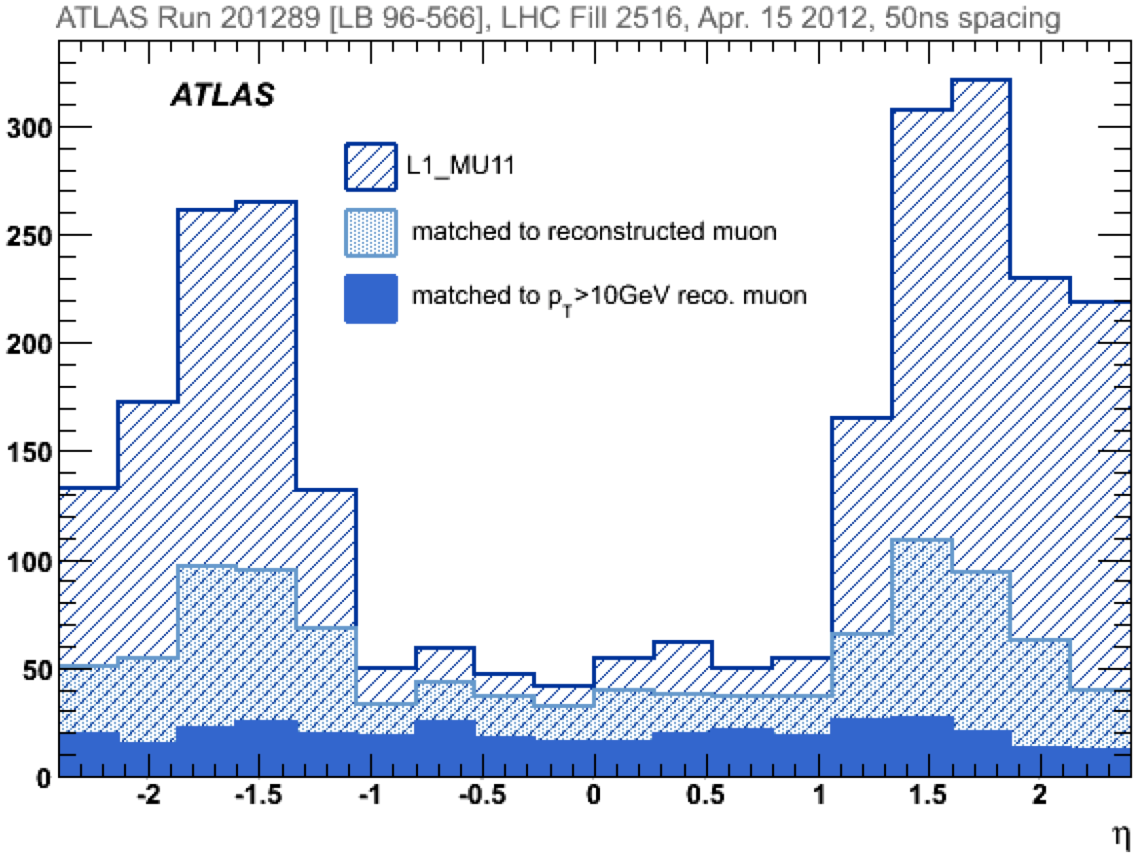}}
\end{center}
\caption{New muon small wheel in $z$-$y$ view (left) and the current Level-1 muon trigger rate in the forward region (right). The hashed area shows the current total (fake+real) rate while the light and dark blue areas show the rate due to real muons.}
\label{fig:smallWheel}
\end{figure}

\begin{figure}[tb]
\begin{center}
\subfigure{\includegraphics[width=0.49\textwidth]{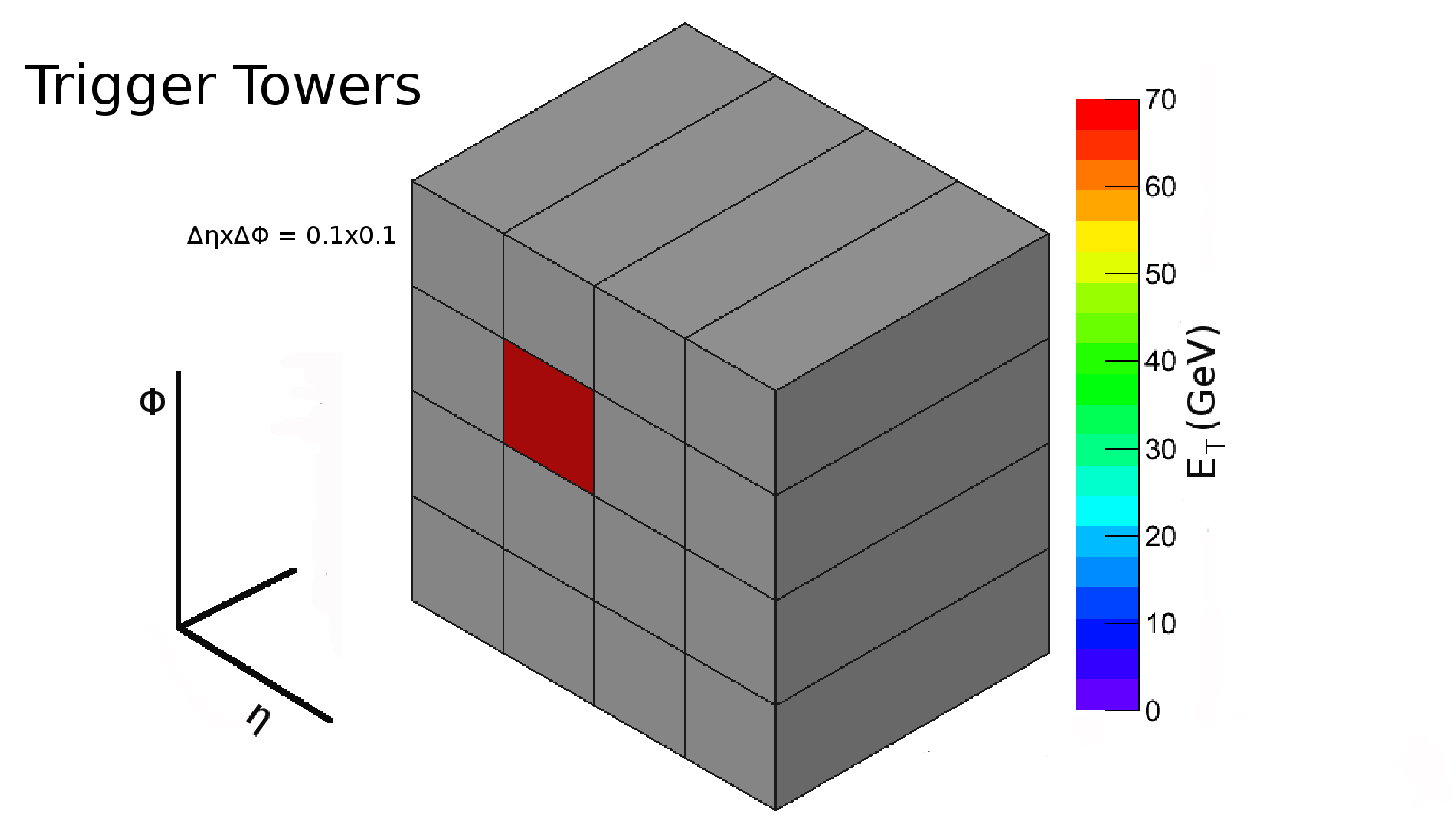}}
\subfigure{\includegraphics[width=0.49\textwidth]{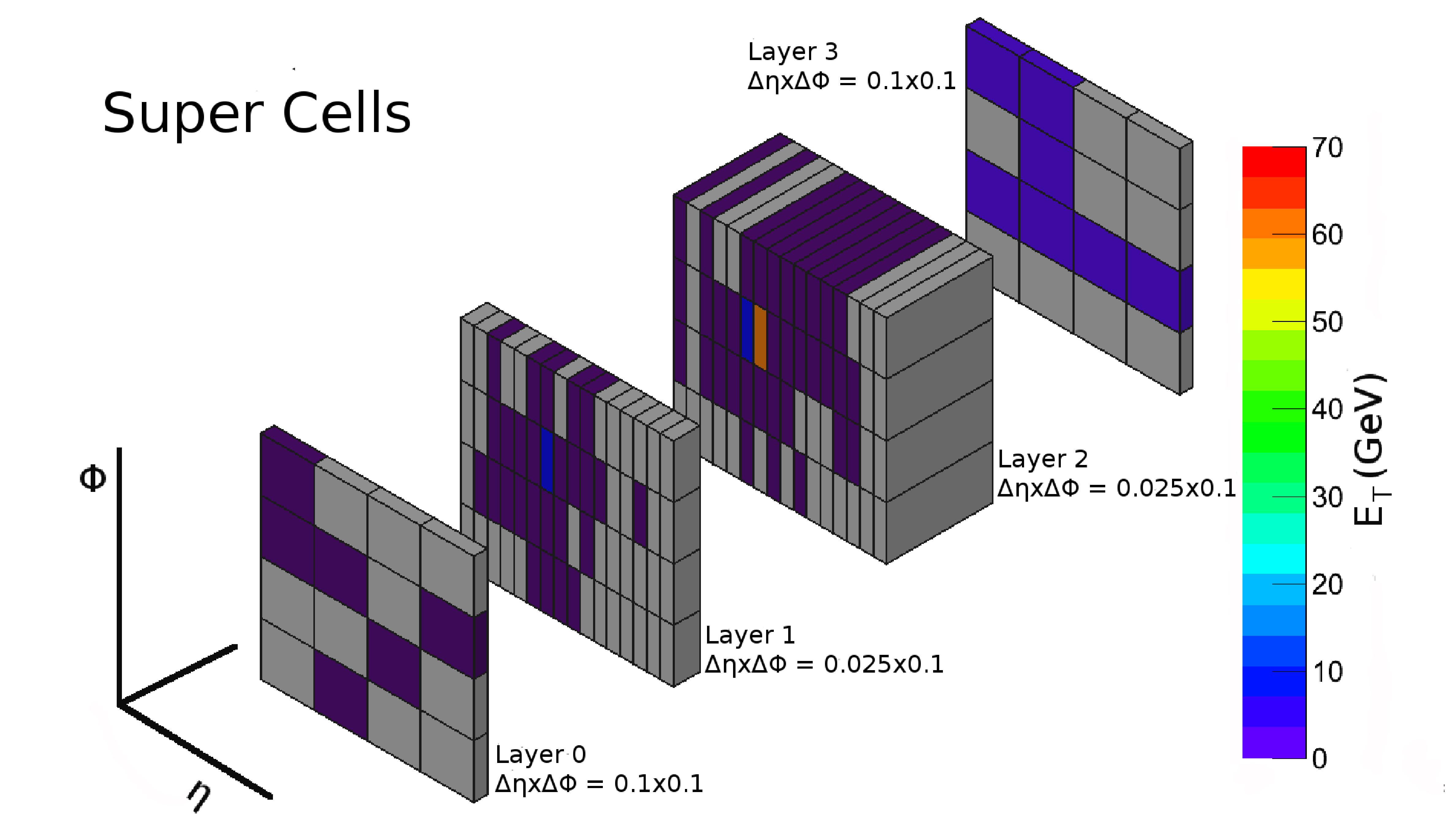}}
\end{center}
\caption{Left: current trigger tower setup with $0.1\times0.1$ towers. Right: super cell layout with finer granularity and layer information.}
\label{fig:LArSCTT}
\end{figure}

\section{Phase-II upgrade}
The LHC upgrade for Phase-II in 2023-2025 increases the instantaneous luminosity to $\sim5 \times10^{34}$ cm$^{-2}$s$^{-1}$. Additionally, by the end of Run 3 the current tracking system would have sustained significant radiation damage. The planned replacement will be an all silicon Inner Tracker (ITk) designed to handle a more than twofold increase in instantaneous luminosity. The inner part of the tracker will be a pixel detector with extended forward tracking. The current outer transition radiation tracker will be replaced by dual layer silicon strip modules (Fig~\ref{fig:itk}). 
The aim of the upgrade is to improve the radiation hardness and forward coverage of the tracker. \par

Further upgrades include the addition of a new hardware trigger resulting in two hardware triggers: level 0 and 1. One of these levels would solely handle tracks while the other trigger uses calorimeter and muon information for trigger decisions. \par

The calorimeter readout will become even more refined with the full calorimeter granularity available at the trigger level. A faster muon trigger is also a part of the Phase-II upgrade which will help maintain reasonable trigger rates in the high luminosity environment.

\begin{figure}[tb]
\begin{center}
\includegraphics[width=0.55\textwidth]{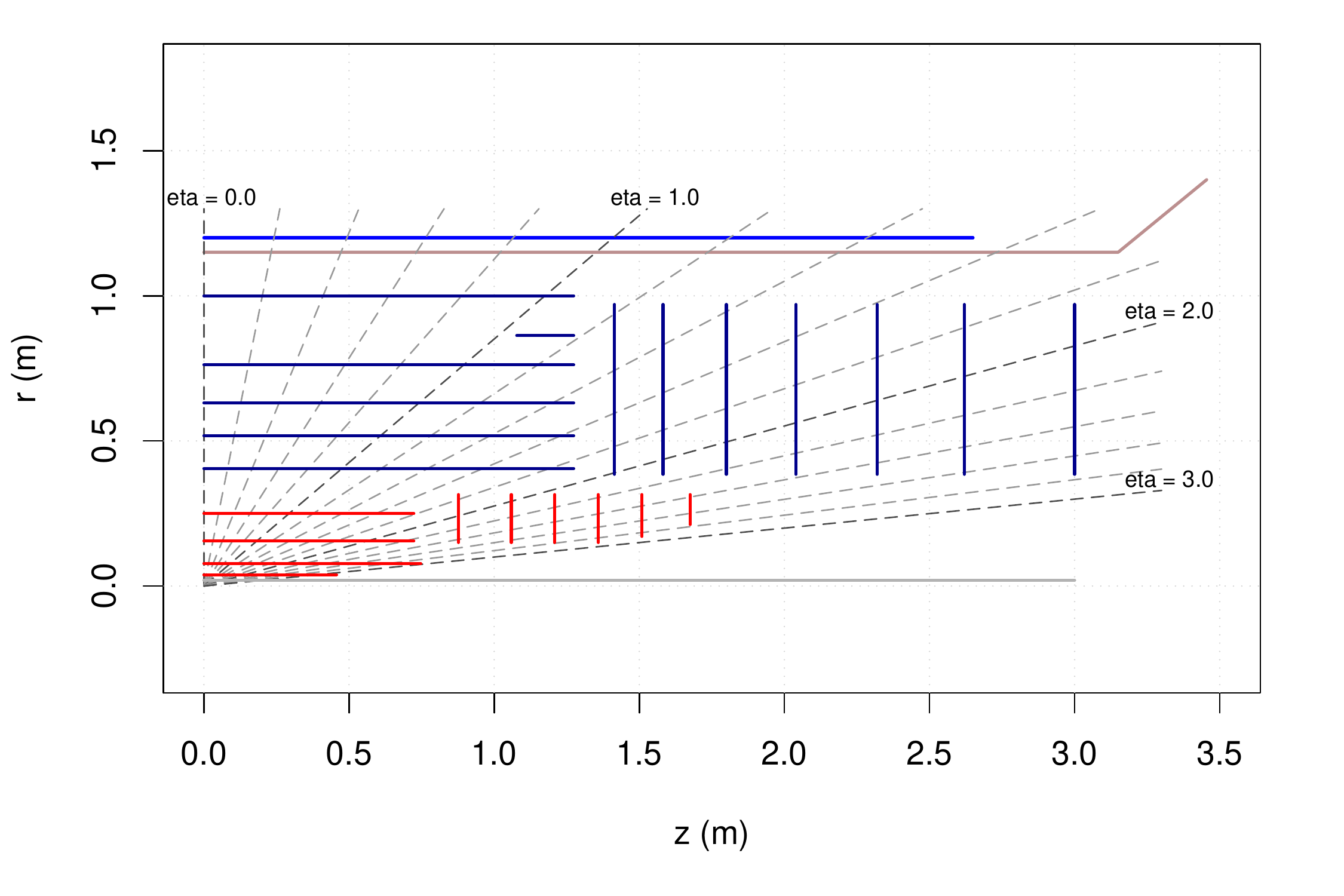}
\end{center}
\caption{ITk layout.}
\label{fig:itk}
\end{figure}

\section{Conclusions}

To continue to increase the precision of our Higgs measurements and probe for new physics with the LHC significant amounts of data are necessary. The proposed upgrade schedule aims at achieving similar performance to Run 1 with large instantaneous luminosity. The upgrades for Phase-I are already at a well defined stage while the Phase-II plans are rapidly developing. With these upgrades the LHC is projected to collect $\sim3000$ \ifb\ over two decades which will be essential in studying physics at the TeV scale.


\end{document}